\documentclass[a4paper,fleqn,usenatbib]{mnras}
\usepackage{amsmath} 
\usepackage{amssymb} 
\usepackage{graphicx}
\usepackage{grffile}
\usepackage[dvips]{epsfig}
\usepackage{epsfig}  
\usepackage{color}
\usepackage{caption}
\usepackage{hyperref}
\usepackage{bm}
\def\be{\begin{equation}}
\def\ee{\end{equation}}
\def\ba{\begin{eqnarray}}
\def\ea{\end{eqnarray}}

\definecolor{red}{rgb}{1,0.0,0.0}

\definecolor{darkgreen}{rgb}{0.0,0.5,0.0}

\newcommand{\tol}{Tololo 1214-277}
\newcommand{\HI}{{\text{H\MakeUppercase{\romannumeral 1}}} }

\newcommand{\lya}{\ifmmode{{\rm Ly}\alpha}\else Ly$\alpha$\ \fi}
\newcommand{\cm}{\ifmmode{{\rm cm}}\else cm\fi}

\newcommand{\ergps}{\,{\rm erg}\,{\rm s}\ifmmode{}^{-1}\else ${}^{-1}$\fi}
\newcommand{\Mpch}{\,{\rm Mpc}\,\ifmmode h^{-1}\else $h^{-1}$\fi}

\newcommand{\kms}{\ifmmode\mathrm{km\ s}^{-1}\else km s$^{-1}$\fi}
\newcommand{\sigmaclump}{$54.3\pm 0.6$ km s$^{-1}$}
\newcommand{\inftyclump}{$54.3\pm 5.1$ km s$^{-1}$}
\newcommand{\probaclump}{$0.96\pm 0.01$}

\begin{document}

%=========================================================================
%		FRONT MATTER
%=========================================================================
\title[An atypical \lya dwarf galaxy]{
Modelling the gas kinematics of an atypical Lyman-alpha emitting compact dwarf galaxy}
\author[J.E. Forero-Romero et al.]
{Jaime E. Forero-Romero $^{1}$ \thanks{je.forero@uniandes.edu.co},
Max Gronke$^2$, 
Maria Camila Remolina-Guti\'errez$^1$,
\newauthor
Nicol\'as Garavito-Camargo$^3$, 
Mark Dijkstra$^2$\\
$^1$ Departamento de F\'isica, Universidad de los Andes, Cra. 1
  No. 18A-10 Edificio Ip, CP 111711, Bogot\'a, Colombia \\
$^2$ Institute of Theoretical Astrophysics, University of Oslo,
Postboks 1029 Blindern, NO-0315 Oslo, Norway.\\
$^3$ Department of Astronomy, University of Arizona, 933 North Cherry
Avenue, Tucson, AZ 85721, USA. 
}

%\vspace*{6pt}\\

%}
\maketitle

\begin{abstract}
	
Star-forming Compact Dwarf Galaxies (CDGs) 
resemble the expected pristine conditions of the first galaxies in the
Universe and are the best systems to test models on primordial galaxy
formation and evolution. 
Here we report on one of such CDGs, \tol, which presents
a broad, single peaked, highly symmetric \lya emission line that had
evaded theoretical interpretation so far.  
In this paper we reproduce for the first time these line features with two
different physically motivated kinematic models: 
an interstellar medium composed by outflowing clumps with 
random motions and an homogeneous gaseous sphere undergoing solid body
rotation.
The multiphase model requires a clump velocity dispersion of
\sigmaclump\ with outflows of \inftyclump\, while the
bulk rotation velocity is constrained to be $348^{+75}_{-48}$ \kms.
We argue that the results from the multiphase model 
provide a correct interpretation of the data.
In that case the clump velocity dispersion implies a dynamical mass of
$2\times 10^{9}$ M$_{\odot}$,  ten times its baryonic mass.
If future kinematic maps of \tol\ confirm the velocities suggested by
the multiphase model, it would provide additional support to expect
such kinematic state in primordial galaxies, opening
the opportunity to use the models and methods presented in this
  paper to constrain the physics of star formation and feedback in the
  early generation of \lya emitting galaxies.
\end{abstract}

\begin{keywords}
galaxies: dwarf --- galaxies: individual:\tol\ --- radiative transfer --- Methods: numerical 
\end{keywords}

%=========================================================================
%		PAPER CONTENT
%=========================================================================

%*************************************************************************
\section{Introduction}
\label{sec:introduction}

Primordial galaxies have not been detected yet. 
However, dwarf star forming galaxies with a low metallicity content
are seen as templates to understand the early galaxy evolution process. 
Over fifty years ago it was realized that young galaxies could be
detected through a strong \lya line emission \citep{PartridgePeebles}.    

This theoretical prediction was only confirmed thirty years later on
distant, relatively young, not primordial, galaxies \citep{1998ApJ...498L..93D}.
Currently Lyman Alpha Emitting galaxies (LAEs) are commonly targeted
in surveys. 
The presence of the \lya emission line gives confirmation of
the distance to a galaxy and provides clues about the stellar
population and inter-stellar medium conditions regulating the
\lya emission.
A careful clustering analysis of LAEs can also yield clues about their link
to dark matter halos
\citep{2004AJ....128.2073H,2007ApJ...671..278G,2007ApJ...668...15K,2008MNRAS.391.1589O,2010MNRAS.409..184P,2013MNRAS.431.1777G,2016ApJ...828....5M}. 

The \lya emission line is not exclusive of distant galaxies. 
There are local Universe surveys that target \lya emission in nearby
dwarf star forming galaxies.
The study of nearby LAE samples has allowed the study of other
indicators that might be more difficult to obtain for distant galaxies
such as galaxy morphology, dust attenuation, neutral hydrogen contents and
ionization state. See \cite{Hayes15} and references therein for a review.

However, the physical interpretation of \lya observations is
not straightforward \citep{LARS,2015ApJ...805...14R}. 
This is due to the resonant nature of the \lya line. 
A \lya photon follows a diffusion-like process before escaping
the galaxy or being absorbed by dust. 
The resulting line profile becomes sensitive to the dynamical, chemical
and thermal conditions in the interstellar medium. 
There are few analytical tools available to interpret the
emerging \lya line
\citep{Harrington73,1991ApJ...370L..85N,LoebRybicki,2006ApJ...649...14D,2006ApJ...645..792T}. 
They are applicable only in highly idealized conditions that
are hardly met in real astrophysical systems. 
For these reasons the interpretation of \lya observations
requires state-of-the-art Monte Carlo radiative transfer simulations.   

Observed \lya line profiles usually present a single peak shifted
redwards from the line's center. 
Sometimes a double peak is present but the asymmetry persists with 
the peak on the red side being stronger \citep[e.g.][]{2010ApJ...717..289S,Erb14,Trainor16}. 
This can be explained as the results of multiple \lya photon
scatterings through a homogeneous medium such as an (outflowing) shell of neutral Hydrogen
\citep{2006A&A...460..397V,Orsi12,2012ApJ...751...29Y,2015ApJ...812..123G}.

\tol\ is a compact dwarf galaxy (CDG) that presents a
strong \lya emission with puzzling 
features: the line is highly symmetric, single peaked and broad 
\citep{Thuan97}.
The existence of this Symmetric Lyman Alpha Emitter (SLAE) raises the question whether some high
redshift LAEs have asymmetric lines because the blue half was
truncated by the intergalactic medium \citep{2007MNRAS.377.1175D}. 
In this case the \lya radiation could emerge as a low surface
brightness glow, which may be connected to \lya halos, while also
influencing the way LAEs can be used as a probe of reionization
\citep[see the review by][and references therein]{2014PASA...31...40D}. 

Attempts to explain the atypical \lya features in \tol\ with
conventional models based on a expanding shell have not been successful
so far \citep{mashesse03,2015A&A...578A...7V}.

Motivated by observations of other CDGs
  that show gas kinematics ranging from pure rotation and low velocity
dispersion to high velocity dispersion without a clear rotation pattern
\citep{2015A&A...577A..21C,2017arXiv170809407C,2017A&A...600A.125C}
we perform here a new study to explain \tol's \lya emission features under two physical
conditions for the interstellar medium: multiphase outflows and pure rotation.

Additional motivation for the outflowing multiphase model \citep[as presented
in][]{Gronke2016} is that some dwarf galaxies are expected to
present outflows.  
Observationally, outflows have been detected in few local dwarf galaxies
\citep{1998ApJ...506..222M,2005MNRAS.358.1453O}. 
Besides, clumpy multiphase outflows are capable to explain \lya features around
star-forming galaxies
\citep{2010ApJ...717..289S,2012MNRAS.424.1672D}.
In addition, due to the cooling properties of gas, multiphase media are expected
in a range of astrophysical systems
\citep[][]{1977ApJ...218..148M}. 

Further  motivation for the model of pure rotation without outflows 
\citep[as presented in][]{GaravitoCamargo2014} is that dwarf galaxies show coherent
rotation features  \citep{2009A&A...493..871S} and it is expected that
some of them have high neutral gas contents with long quiescent phases
without high gas outflows triggered by supernova activity
\citep{2005A&A...433L...1B,2008ApJ...672..888T,2013MNRAS.434.2491G}.

The models we use correspond to simplified geometrical configurations.
This allows us to perform a deep exploration of parameter space and
gain some physical insight into \tol's kinematic properties.
In this paper we show, for the first time in the literature, that \tol's
\lya profile can be explicitly modeled by either of these two
models. 

In the next section we review the observational characteristics of
\tol, then we summarize the main features in the multiphase and
rotation models to explain how we fit their free parameters 
to the \tol's \lya line shape.
We use the results to interpret them in terms of the galaxy's
dynamical mass and argue why in this case the multiphase model should
be preferred over the rotation model.

\section{Observations}

\begin{table}
\begin{center}
\begin{tabular}{lr}\hline
$\alpha$(2000)& 12h17min17.1s\\
$\delta$(2000)& -28d02m32s\\
$l$, $b$ (deg) & 294, 34\\
$m_V$ & 17.5\\
  M$_V$ & -17.6\\ 
Redshift & $0.026\pm0.001$ \\
2D half-light radius  & $1.5\pm 0.1$ kpc\\\hline
\end{tabular}
\end{center}
\caption{Basic observational characteristics of \tol\ 
  \citep{Thuan97}.
The 2D half-light radius is computed from the results reported by
\citet{2003A&A...410..481N}.  
\label{obstable}}
\end{table}

\begin{table*}
  \centering
\begin{tabular}{l|lrrl}
\hline
  Parameter  & Description & Fiducial value & Allowed range & Units \\ \hline\hline
$v_{\infty,\,{\rm cl}}$ & Radial cloud velocity & $100.0$ & [$0.0$, $800.0$] & $\,{\rm km\,s}^{-1}$ \\
$\sigma_{\rm {\rm cl}}$ & Random cloud motion &$40.0$ & [$5.0$, $100.0$] & $\,{\rm km\,s}^{-1}$ \\
$P_{\rm {\rm cl}}$ & Probability to be emitted in cloud & $0.35$ & [$0.0$, $1.0$] &  \\
$r_{\rm {\rm cl}}$ & Cloud radius & $100.0$ & [$30.0$, $200.0$] & $\,{\rm pc}$ \\
$H_{\rm em}$ & Emission scale radius & $1000.0$ & [$500.0$, $3.0\times 10^{3}$] & $\,{\rm pc}$ \\
$f_{\rm {\rm cl}}$ & Cloud covering factor & $3.5$ & [$0.8$, $8.0$] &  \\
$T_{\rm {\rm ICM}}$${}^{\dagger}$ & Temperature of ICM & $10^{6}$ & [$3.0\times 10^{5}$, $5.0\times 10^{7}$] & $\,{\rm K}$ \\
$n_{\rm HI,\,{\rm ICM}}$${}^{\dagger}$ & \HI number density in ICM & $5.0\times 10^{-8}$ & [$10^{-12}$, $10^{-6}$] & $\,{\rm cm}^{-3}$ \\
$\sigma_{\rm i}$ & Width of emission profile & $50.0$ & [$5.0$, $100.0$] & $\,{\rm km\,s}^{-1}$ \\
$T_{\rm {\rm cl}}$${}^{\dagger}$ & Temperature in clouds & $10^{4}$ & [$5.0\times 10^{3}$, $5.0\times 10^{4}$] & $\,{\rm K}$ \\
$\beta_{\rm cl}$ & Steepness of the radial velocity profile & $1.5$ & [$1.1$, $2.5$] &  \\
$\tilde\sigma_{\rm d, cl}$${}^{\dagger}$ & Dust content in clumps &
  $3.2\times10^{-22}$ & [$4.7\times 10^{-24}$, $1.6\times 10^{-21}$] &
  $\,\cm^{2}$ \\
$\zeta_d$${}^{\dagger}$ & Ratio of ICM to cloud dust abundance &
  $0.01$ & [$10^{-4}$, $0.1$] & \\
$n_{\rm HI,\,{\rm cl}}$${}^{\dagger}$ & \HI number density in
  clouds & $0.35$ & [$0.03$, $3.0$] & $\,{\rm cm}^{-3}$ \\
\hline
\end{tabular}
  \caption{Overview of the parameters in the multiphase model and its fiducial
    values. Variables marked with ${}^{\dagger}$ were drawn in
    log-space. Table reproduced from \citet{Gronke2016}.}
  \label{tab:models}
\end{table*}

\tol's basic observational characteristics are summarized in Table \ref{obstable}.
Its receding velocity is $7785\pm 50$km s$^{-1}$ translates
into a distance of $106.6$ Mpc, using a Hubble constant value of $H_{0}$=73
km s$^{-1}$ Mpc$^{-1}$.

Archival Chandra X-ray data\footnote{\url{http://cxc.harvard.edu/cda/}} does not show any
detection for \tol. 
This lack of detection motivates our assumption that \lya emission is
powered by star formation only.

The observed flux for the \lya line is $\sim
8.1\times 10^{-14}$ erg cm$^{-2}$ s$^{-1}$ \citep{Thuan97}.
The \lya Equivalent Width is $70$\AA\ and its H$\beta$ flux is 
$1.62\times 10^{-14}$ erg cm$^{-2}$ s$^{-1}$ \AA$^{-1}$
which gives a \lya/H$\beta$ flux ratio of
4.9$\pm$0.1 \citep{Izotov04}.
Comparing the \lya/H$\beta$ ratio with the theoretical
expectation from case B recombination of $23.3$ \citep{Hummer1987} one
can estimate an escape fraction of $20$\% for \lya radiation.
Figure \ref{fig:results} shows \tol's \lya\ profile reported by
\cite{mashesse03}. This measurement was made with the Space Telescope
Imaging Spectrograph on board the Hubble Space Telescope, with a
spectral resolution of $\sim 37$\kms\ at the  \lya\ wavelength.  

The \lya flux values correspond to a luminosity of
$L_{Ly\alpha}=2.2\times 10^{42}$ erg s$^{-1}$, which in turn
translates  into a lower bound for the star formation rate of $2.0$
M$_{\odot}$ yr$^{-1}$ after using a standard conversion factor between
luminosity and star formation rate of $9.1\times 10^{-43}$
$L_{\rm Ly\alpha} / (\mathrm{erg}\,\mathrm{s}^{-1})$ M$_{\odot}$ yr$^{-1}$ 
without any
correction by extinction
\citep{Kennicutt98}.

The bolometric UV luminosity is $9.43\pm1.94 \times 10^{8}$
L$_{\odot}$ as measured by GALEX. Without any correction by extinction
and following the empirical relation by \cite{Kennicutt98}, this
corresponds to a star formation rate of $0.35\pm 0.05$ M$_{\odot}$
yr$^{-1}$. 
The absolute magnitude in the $V$ band translates into a luminosity of
$8.9\times 10^{8}$ L$_{\odot}$.  
Its metallicity is $\sim Z_{\odot}/24$ as derived from
optical spectroscopy \citep{Izotov04}. 
% using http://tomdwelly.com/tools_fluxtolum.php
%

The near-infrared fluxes at $3.6$ $\mu$m and $4.5$ $\mu$m are
$7.71\pm0.55\times 10^{-5}$ Jy and $7.98\pm0.71\times 10^{-5}$ Jy
\citep{2008ApJ...678..804E}.
Using the conversion between fluxes and
stellar mass, $M_{\star} =
10^{5.65} \times F_{3.6}^{2.85} \times F_{4.5}^{-1.85} \times
(D/0.05)^2 M_{\odot}$,  calibrated on the Large Magellanic Cloud 
and  where fluxes are in Jy and $D$ is the luminosity
distance to the source in Mpc, we find $M_{\star} = 1.45\pm0.45\times 10^{8}
M_{\odot}$, with a $30\%$ uncertainty coming from the calibration
process \citep{2012AJ....143..139E}.  
There is an upper limit for the  $21$ cm line integrated flux of $<0.10$
Jy km s$^{-1}$  which translates into a upper limit for the neutral
Hydrogen mass of $M<2.65\times 10^{8}$ M$_{\odot}$
\citep{pustilnikmartin07}. 

We compute the projected half-light radius to be $R_s=1.5\pm0.1$ kpc 
from the surface brightness profiles reported by \cite{2003A&A...410..481N}. 
Assuming spherical geometry, one can translate this value into a 3D
half-light radius of $r_s=3R_s/2=2.25$ kpc.
Imaging observations by \cite{2001AJ....121..169F} show that \tol\ is
an isolated field galaxy not belonging to a group or cluster.

\section{Theoretical Models and Parameter Estimation}

\subsection{Multiphase ISM} 

\begin{figure*}
\begin{center}
\includegraphics[width=0.55\textwidth]{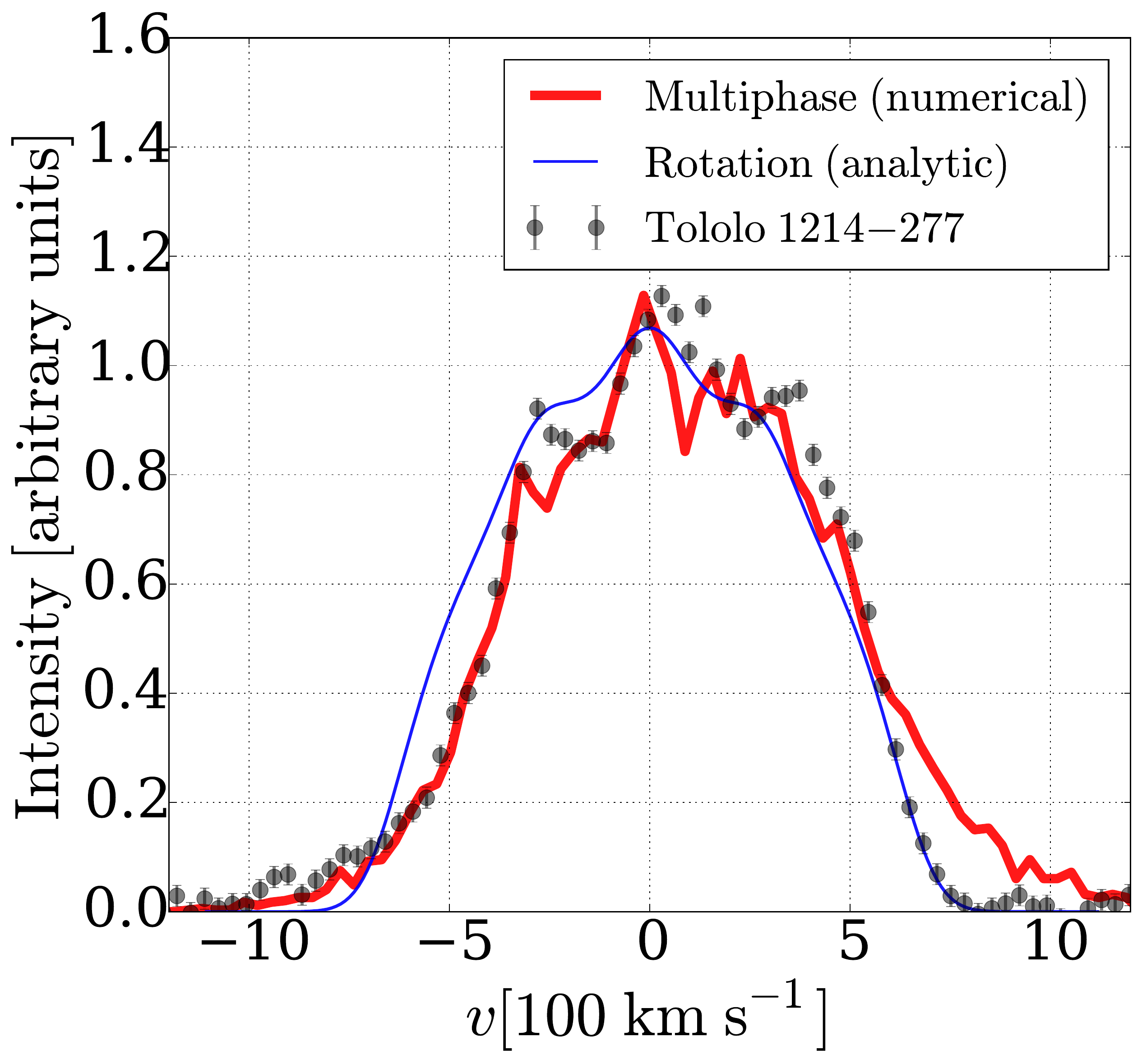}
\caption{Broad, single peaked and highly symmetric \lya emission of \tol.
  Dots correspond to the observational data \citep{mashesse03}. 
The thick red and thin blue curves represent our best fit
to the data using the full radiative transfer simulation with a
rotation and multiphase model, respectively. 
These two different models are able to reproduce the main morphological
features of \tol's \lya line emission.\label{fig:results}}
\end{center}
\end{figure*}

The idealized multiphase model consists of spherical, cold, dense
clumps of neutral hydrogen and dust embedded in a hot, ionized
medium \citep{Gronke2016}. 
The clumps also have a random and an outflowing velocity
component which totals the number of parameters describing the model
to be $14$.
We do not explore inflowing clumps given the slight line asymmetry
redwards to the line center (see the dots in Figure \ref{fig:results})
and thus set $v_{\infty,\,{\rm cl}}>0$.
The parameter description list is in Table \ref{tab:models}.

In order to map out this large parameter space, we randomly drew
$2500$ sets of parameters within an observationally realistic range,
based on the considerations of \citet{Laursen2013ApJ...766..124L}, 
yielding a large variety of single-, double- and triple-peaked
spectra. 
The full analysis of the spectral features as well as
more details on the radiative transfer are presented by
\citet{Gronke2016}.    

We compare each resulting spectra to the observational results from
\tol\ after normalizing the observed and simulated spectra to have a 
flux integral of one.
We build a $\chi^2$ on the normalized flux measurements for each one
of the $2500$ models as

\begin{equation}
\chi^2 = \sum_{i} \frac{({f}_i - \hat{f}_i)^2}{\sigma_i^2}, 
\label{eq:chi2}
\end{equation}
where $i$ iterates over velocity bins, $f_{i}$ is the observed flux,
$\sigma_i$ is the observed flux uncertainty and $\hat{f}_i$ is the
model estimated flux.
As we do not have an analytic expression for $\hat{f}$;
we obtain $\hat{f}$ from the binned results of the Monte Carlo radiative transfer simulations.

We select for further analysis the best $1\%$ models according to the
lowest $\chi^2$ values.
We note that the difference between the lowest and highest $\chi^2$ values in
those $25$ models is close to $\Delta\chi^2 = 3000$, the lowest
$\chi^2$ being close to $\chi^2_{\mathrm{min}}=1200$. 

We run a Kolmogorov-Smirnov (K-S) test to compare each parameter
distribution in the best $25$ models against the parent distribution
of $2500$ models. 
If we obtain a p-value $<0.05$ we conclude that
this parameter can be constrained from the observations, as the distribution for
the best $\chi^2$ models is statistically different to the
distribution from the global sample of $2500$ models.  
In the Appendix \ref{appendix} we complement this analysis using a
random forest classifier to find the most important parameters in
selecting a low $\chi^2$ model.

We finally compute the best values for the constrained parameters as
the values that produce the minimum $\chi^2$.  
We estimate the 1-$\sigma$ uncertainty from a parabolic fit to the
$\chi^2$ as a function of the best constrained parameters around its
corresponding minimum.

\subsection{Bulk Rotation}

The rotation model corresponds to the work presented by
\cite{GaravitoCamargo2014} based on the Monte Carlo code
\texttt{CLARA} \citep{CLARA}. 
In that model the \lya photons are propagated 
within a spherical and homogeneous cloud of HI gas undergoing solid
body rotation.
The sphere is fully characterized by three parameters: the HI line's
center optical  depth $\tau$ measured from the center to its surface, the HI
temperature $T$, and the linear surface velocity $V_{\rm max}$.  
The observed line profile also depends on $\theta$,  the angle between the plane
perpendicular to the rotation axis and the observational
line-of-sight.  
In this model, dust only changes the overall line
normalization and only weakly its shape, i.e. 
dust cannot change the line symmetry or induce a change in the number
of line peaks, moreover it does not change the line width by more than
$1\%$ ($5$ \kms\ in the case of \tol), an effect too small to be
observed at the resolution at which we have \tol's line profile and
also negligible compared to the influence of the other free parameters in the model.
For these reasons we do not include any dust model. 

We use an analytical approximation that captures the most important
effects of rotation onto the \lya line. 
We defer the reader to \citet{GaravitoCamargo2014} for complete
details on the explicit form of this approximation.
To fully explore the parameter space we perform Markov Chain Monte
Carlo (MCMC) calculations with the \texttt{emcee} Python library
\citep{2013PASP..125..306F}. \texttt{emcee} is an open source
optimized implementation of the affine-invariant MCMC sampler
\citep{goodman2010ensemble}.  
The algorithm creates a number of walkers that,
during a sufficient number of steps, generate parameters' combinations
for a specific model.
For each timestep, the code calculates the likelihood of the
combination with respect to the observational data.
The walkers explore the parameter space sampling the Gaussian likelihood
function built as $\propto \exp(-\chi^2/2)$, where the $\chi^2$ follows
the definition in Equation \ref{eq:chi2}. 
We do not have a closed analytic expression for $\hat{f}$, we compute it
by numerical integration of the analytical approximations found in
\citet{GaravitoCamargo2014}.  

We explore flat priors on four parameters: $200<V_{\rm
  max}/\mathrm{km\ s}^{-1} <600$,   $6.0<\log_{10}\tau<9.0$,
$4.0<\log_{10} T/10^4\mathrm{K}< 4.5$ and $0<\theta<90$ using $500$
steps with $24$ walkers for a total of $12000$ points in the chain.
Previous exploratory work shows that it is impossible to fit the
observed line outside these priors.
Finally, we estimate the parameter values from the 16th, 50th and 84th
percentiles.

\begin{figure*}
\begin{center}
  \includegraphics[width=0.6\textwidth]{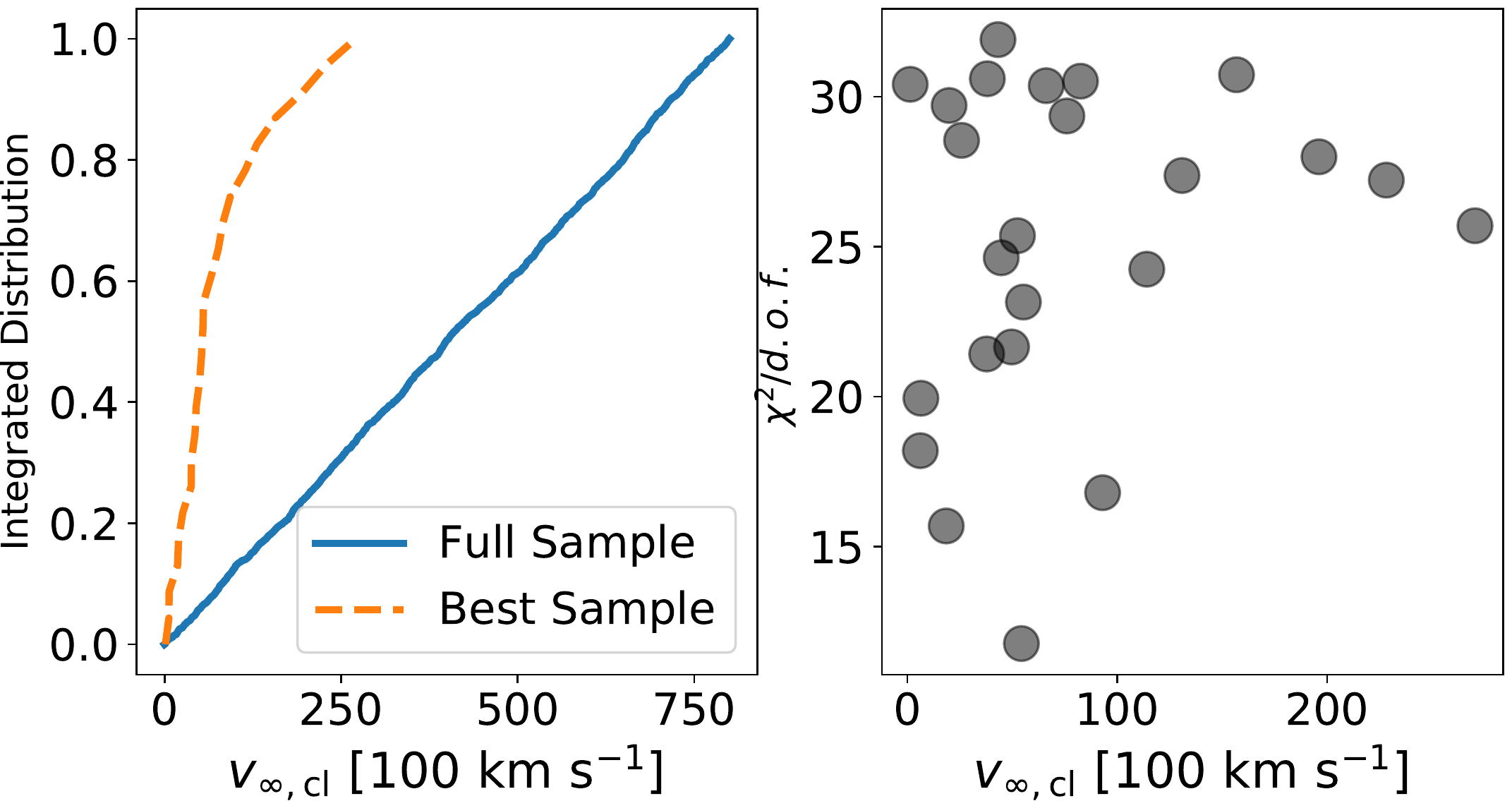}
  \includegraphics[width=0.6\textwidth]{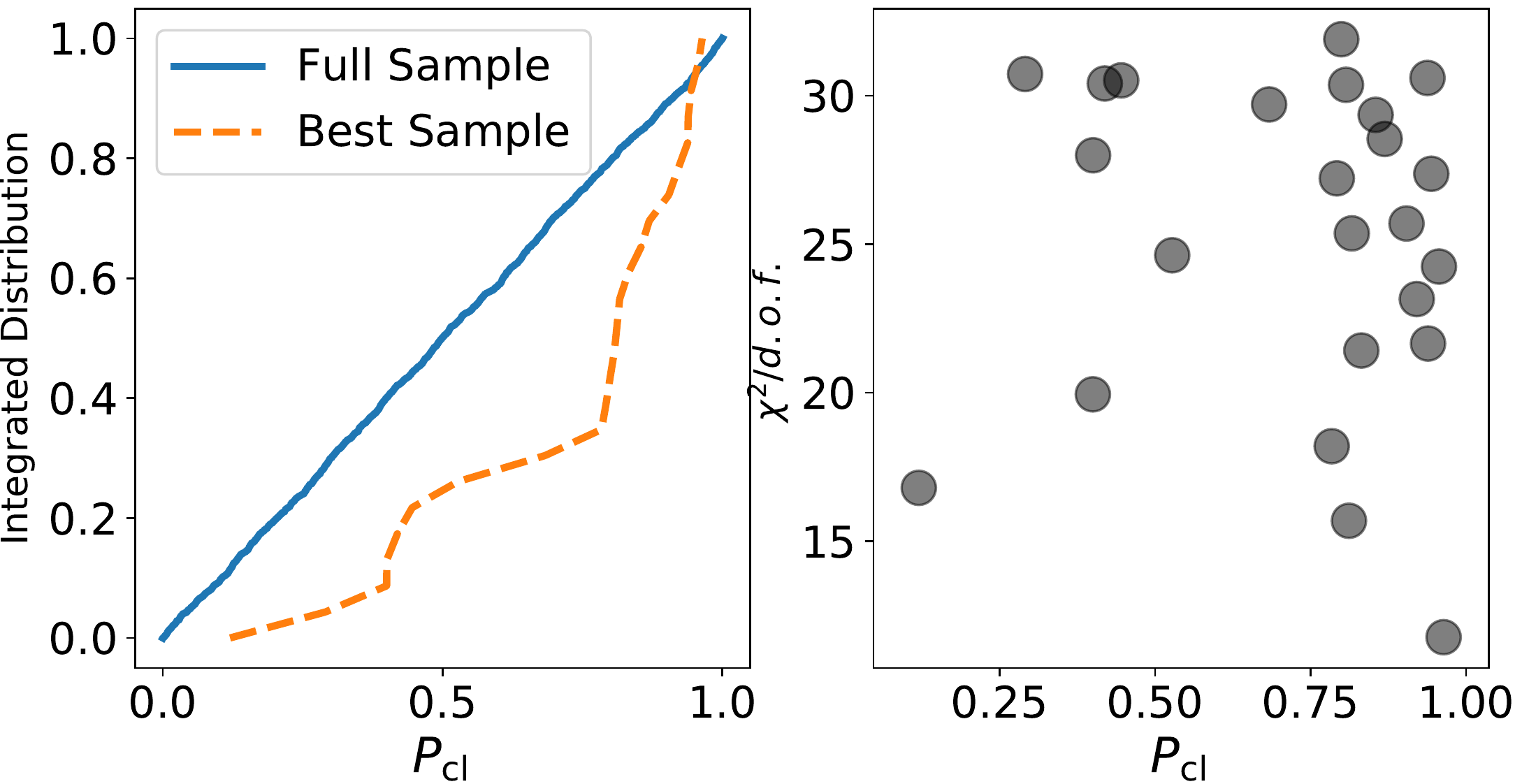}
  \includegraphics[width=0.6\textwidth]{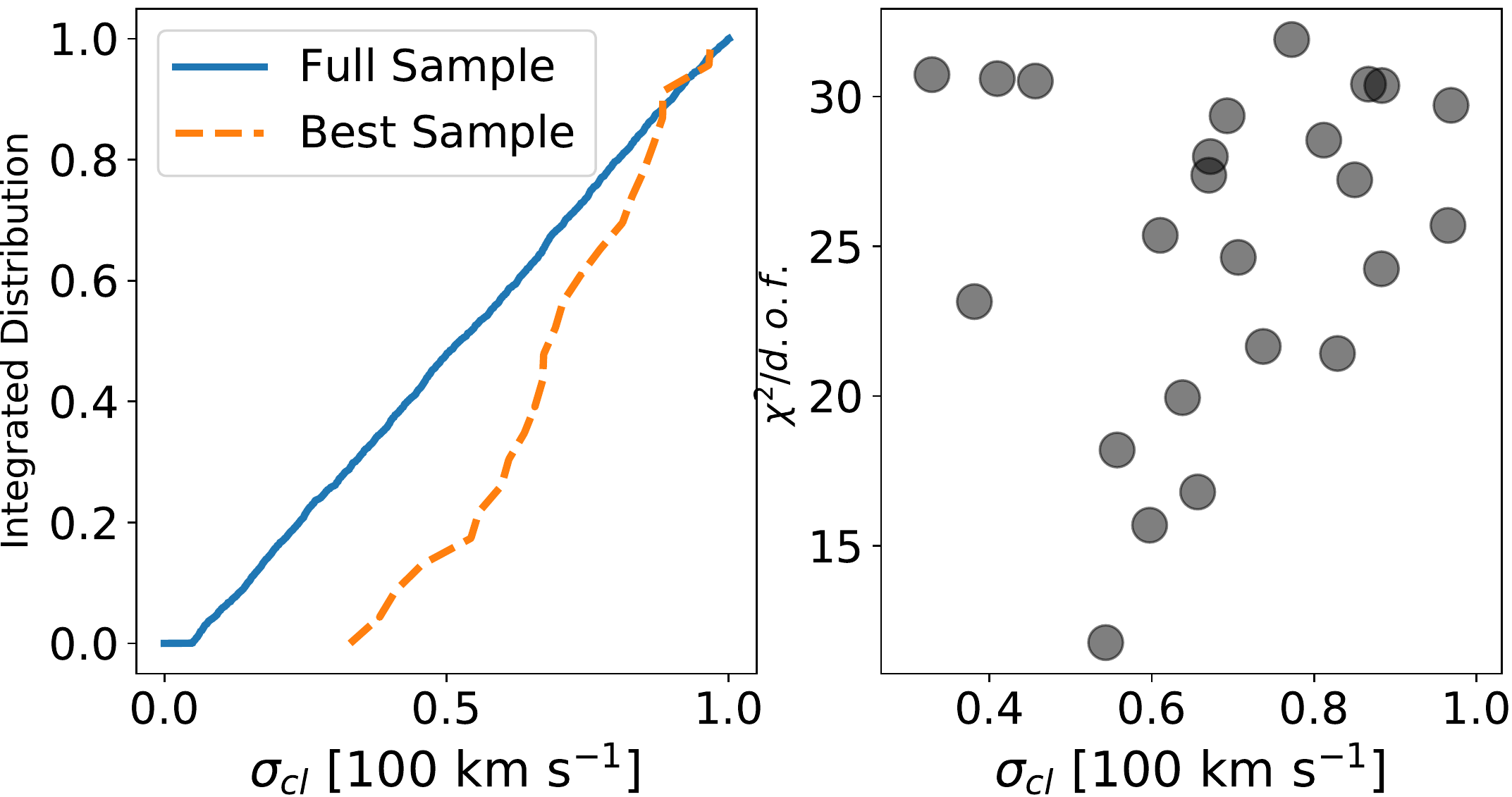}
\caption{Results from the multiphase  model.
  We only show results for the three parameters with a significant
  statistical difference between the models with the lowest $\chi^2/d.o.f.$
  values and its prior distribution. 
  These three parameters are the clump radial velocity at infinity 
  $v_{\infty, \rm{cl}}$; the probability that  the \lya photons were
  emitted in the clumps, $P_{\rm cl}$; the clump velocity dispersion,
  $\sigma_{\rm cl}$.
  The left column corresponds to the parameter's integrated distributions for
  models with the lowest $\chi^2/d.o.f.$ values (dotted line) compared against the
  parameter's prior integrated distributions (continuous line).
  The right column shows a scatter plot of $\chi^2/d.o.f.$
  and its corresponding physical parameters.
  The clump kinematic state is the key ingredient to
  reproduce \tol's \lya line morphology.
  \label{multiphaseresults}
}  
\end{center}
\end{figure*}

\begin{figure*}
\begin{center}
\includegraphics[width=0.8\textwidth]{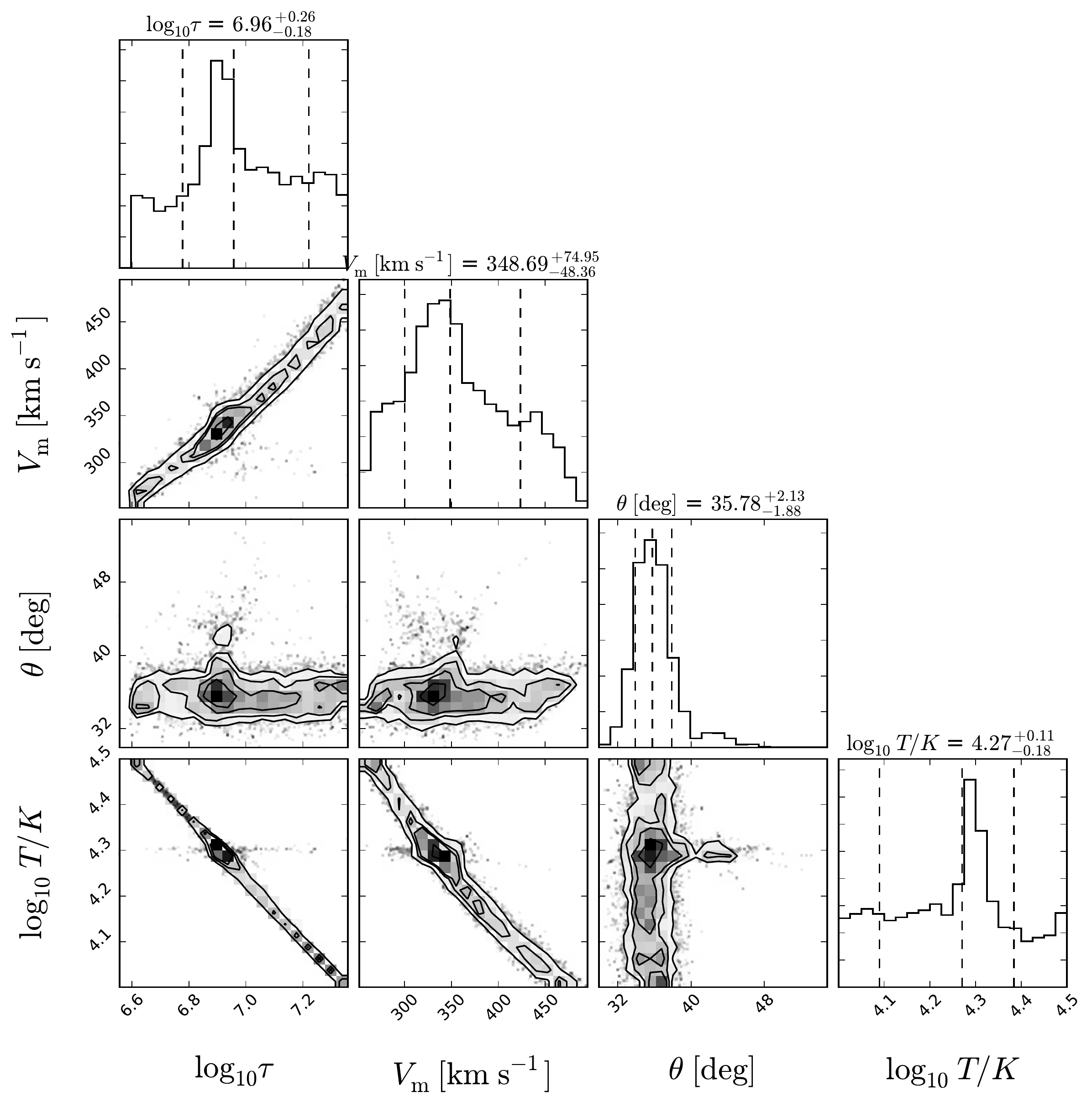}
\caption{Results from the Markov Chain Monte Carlo computation for
    the rotation model. 
    The gray scale indicates the point density in parameter space. 
    The dotted vertical lines in the histograms in the diagonal panels
    represent the 16th, 50th and 84th percentiles. 
    All parameters in the rotation model can be successfully constrained by \tol's
    \lya line morphology to the values written in the top region of
    the diagonal panels.
    \label{emceeresults}} 
\end{center}
\end{figure*}

\section{Results}

Figure 1. Summarizes our main finding.
Dots represent the observational data for \tol\ with the
overplot from our best fits from the analytical solution for the
multiphase model (thick line)  and the rotating homogeneous gas sphere
(thin line). 
The fit to the observations is not perfect. 
However, in spite of the
simplicity of our models, this is the first time that the main
features of this SLAE can be reproduced: a broad, highly symmetric,
single-peaked \lya line. 

This result does not demonstrate that the kinematic features we
include in our models are necessary to reproduce \tol's features, but
at least they show that they are a sufficient condition.
This is a significant step forward to understand the influence that
different kinematics have in producing the atypical line profile shown
by \tol.

In what follows we summarize the values of the kinematic parameters
that managed to explain \tol's \lya profile.

\subsection{Multiphase ISM}

With $14$ free parameters our first concern is discovering which 
parameters matter the most.
From the K-S tests we find that only 3 parameters 
are confidently constrained by \tol's line shape;
$v_{\infty,\rm{cl}}$ (p-value  $10^{-18}$), $\sigma_{\rm cl}$ (p-value
$10^{-4}$) and $P_{\rm cl}$ (p-value $10^{-4}$). 

The low p-values are illustrated by the results shown in Figure \ref{multiphaseresults}.
Left column shows the difference between the integrated distributions
of the full sample (2500 input models) and the $1\%$ models with the lowest
$\chi^2/d.o.f.$, where we use the total number of degrees of freedom,
$d.o.f.=104$. 
The right column shows the actual $\chi^2/d.o.f.$ 
and its corresponding parameter value. 
Under these conditions we find $\sigma_{\rm cl}=$\sigmaclump,
$v_{\infty {\rm, cl}}=$\inftyclump\ and $P_{\rm cl}=$\probaclump, 
  with the minimum $\chi^2/d.o.f.=11.7$. This relatively high value
    for the reduced $\chi^2$ could be interpreted as low statistical
    significance. 
However, we stress once more that the is the first time that the main
features of \tol\ can be reproduced, making this model an useful tool
to guide the interpretation of complex observational data. 
These results certainly open up the path to future searches to
construct new models.

The results from this model can be qualitatively explained as follows.
Due to the large fraction of \lya photons being emitted within the
moving clumps ($P_{\rm cl}\sim 1$) the `intrinsic' profile closely follows 
the clump kinematics. 
In other words, the width of the intrinsic line
is set by $\sigma_{\mathrm{cl}}$ and its median offset mainly by
$v_{\infty,{\rm cl}}$ and $\beta_{\rm cl}$, the exponent defining
  the steepness of the radial velocity profile. 
Furthermore, the relatively low mean number of clumps per line of
sight, $0.8<f_{\mathrm{cl}}<8$, combined with the high velocity dispersion
of the clumps implies existence of low-density inter-clump 
regions where \lya photons can freely propagate.
This gives result to an emergent spectrum close to the intrinsic one,
explaining the high flux at the line's center.    

Because the width of the observable spectrum is hence set primarily by
$\sigma_{\rm cl}$, a lower velocity dispersion would produce a
narrower line and thus a worse fit. 
From the lower right panel in Figure \ref{multiphaseresults} we find
that in fact it is unlikely that the clump velocity dispersion is
lower than $50$ \kms. 

Having constrained only 3 parameters one might wonder why the other 11
parameters do not seem to matter.
In our case this can be explained by the large value of $P_{\rm cl}$
favored by \tol's observations. 
A large probability of having \lya\ photons emitted in the clumps
makes radiative transfer effects, and therefore other parameters such
as the clump column density, less relevant for the emergent line
profile.

However, we cannot discard that another region of parameter space
could also yield a good fit to the observed line-profile.
This might be interesting to explore in the
future with, e.g., a higher sampling of the parameter space once new 
observations  yield more details on \tol's kinematic structure. 

\subsection{Bulk Rotation}

The results for this model are easier to interpret due to the fewer
number of free parameters and its explicit influence on the
semi-analytic solution.
The results are summarized in  Figure \ref{emceeresults}. 
From this analysis we find that the best parameters in the rotation
model are a rotational velocity of  $V_{\rm max}=348^{+75}_{-48}$
\kms, a neutral Hydrogen optical depth of
$\log_{10}\tau=6.96^{+0.26}_{-0.18}$,  and an inter-stellar medium
temperature of $\log_{10} T/\mathrm {K} = 4.27^{+0.11}_{-0.18}$.   
We are also able to constrain the angle between the plane
perpendicular to the rotation axis and the observational line-of-sight
to $\theta = 35.78^{+2.13}_{-1.88}$ degrees.
This model cannot reproduce the slight asymmetry present in \tol's
\lya line; most probably this would require an small amount of
outflows, a feature that is not present in the model provided by
\cite{GaravitoCamargo2014}. 

The preferred value for the rotational velocity can be explained as
follows. 
Lower rotational velocities than the favored value would produce a
double peaked line as the different doppler shifts produced on
different regions of the rotating sphere would not be large enough to
smear the double peaks into a single peak
\citep{GaravitoCamargo2014}. 
For the same reason, higher rotation velocities could produce a single
peak but the line would be broader than it is observed.
The fact that the velocity and optical depth priors were wide enough,
allows us to suggest that the current values for the spherical model
found by the MCMC are robust given the observational constraints.

\section{Discussion}

\tol\ presents a \lya emission line with puzzling features.
Its flux at the line's center is high compared to other LAEs at low
redshift and the broad, highly symmetrical peak.
SLAEs are virtually absent from other LAE surveys at low and high redshift
\citep{2012ApJ...751...29Y,LARS,Erb14,Trainor16}. 
Simple shell models
fail to reproduce such a spectrum as reported by \cite{2015A&A...578A...7V}.  
In the previous sections we show how these characteristics can be
explained by two different kinematic models: multiphase ISM and solid
body rotation.

Which model could be closer to the actual kinematic conditions in \tol?  
Integral Field Unit (IFU) observations of other CDGs seem to
favor the multiphase model \citep{2015A&A...577A..21C,2017arXiv170809407C,2017A&A...600A.125C}.
These observations  were performed with the Visible Multi-Object Spectrograph (VIMOS)
\citep{2003SPIE.4841.1670L}.
The spatial sampling was $0.67^{\prime\prime}$ and covered about
$30^{\prime\prime}\times 30^{\prime\prime}$ on the sky. 
They observed nine Blue Compact Galaxies (BCGs) to produce two-dimensional maps of the
continuum and prominent emission lines to finally compute velocity
field maps using the H$\alpha$ emission line. 
They find velocity fields ranging from simple rotation patterns 
(with amplitudes of $10-120$ \kms) to highly irregular. 
The typical velocity dispersion values are in the range $10-50$ \kms
with the exception of one galaxy that shows a dispersion of $130$
\kms. 

These results disfavor the high rotational velocity of $V_{m}\sim350$
\kms\  that we find in the pure rotation model. 
On the other hand the results from the multiphase model yield a
velocity dispersion $\sigma_{\rm cl}=$\sigmaclump, consistent with
other observations.
We can now estimate a value for the dynamical mass using the
constraints for the velocity dispersion, $\sigma$,  in a spherical
system localized in a region of size $r$     

\begin{equation}
M_{\rm dyn} = 3 \frac{\sigma ^{2}r}{G} = 3.48\times10^{9}
\left(\frac{\sigma}{100\ \mathrm{km\ s}^{-1}}\right)^2\left(\frac{r}{\mathrm{kpc}}\right)
M_{\odot}. 
\end{equation}

Assuming that the \lya emission is entirely powered by star formation 
we use the 3D half-light radius $r_s=2.25$ kpc as the typical size
for the HI region. 
With $\sigma=\sigma_{\rm{cl}}=$\sigmaclump we obtain a dynamical mass
of  $M_{\rm dyn}=2.31\pm0.04 \times 10^{9}$ $M_{\odot}$, which is ten
times the estimated baryonic mass in \tol. 

\begin{figure}
\begin{center}
\includegraphics[width=0.46\textwidth]{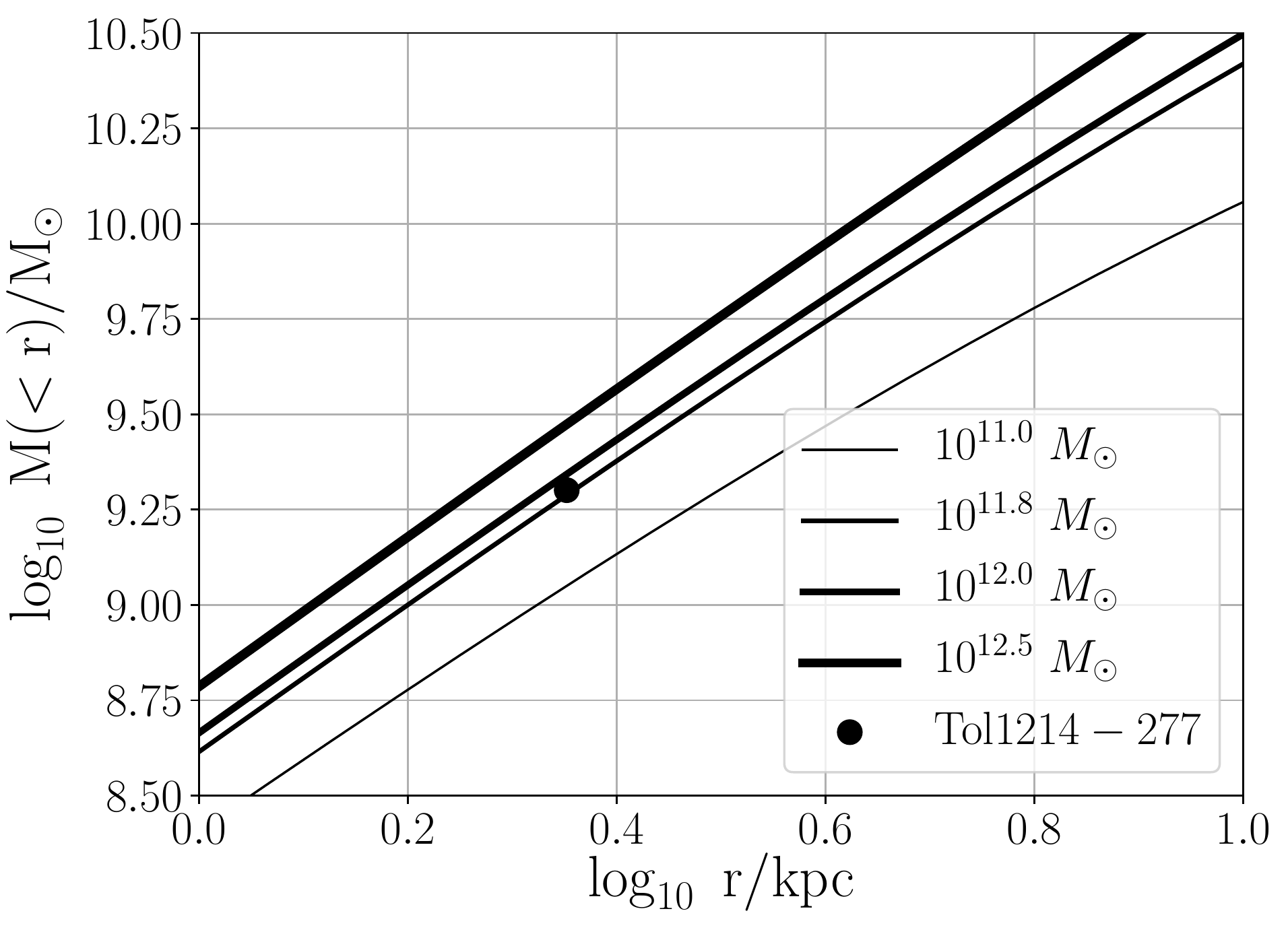}
\caption{Enclosed mass in a spherical system as a function of radius.
  Lines correspond to the expectation for dark matter halos following
  an NFW profile with
  different virial masses as shown in the legend. The circle corresponds
  to the dynamical mass estimates for \tol.  
  Under this conditions the dynamical mass estimates for \tol\ are consistent with the
  galaxy being hosted by a dark matter halo of
  $\sim 6\times10^{11}$ M$_{\odot}$ in mass.
    \label{fig:mass}} 
\end{center}
\end{figure}

A larger dynamical mass estimate over its baryonic mass hints that
\tol\ is dark matter dominated.  
Following the methodology of \citet{2011ApJ...726..108T} we estimate
that a dark matter halo with a virial mass $\sim 6\times 10^{11}$
M$_{\odot}$ and a virial radius $220$ kpc could explain this dynamical mass.
This estimate is based on computing the integrated mass profile of a
spherical dark matter halo with a Navarro-Frenk-White profile with its
concentration following the median mass-concentration
relationship found in the Bolshoi simulation
\citep{2012MNRAS.423.3018P,2016ApJ...832..169P}.  
Figure \ref{fig:mass} shows the enclosed mass as a function of radius
for different dark matter halos together with the \tol's dynamical
mass estimate. 
This should be considered as an upper bound as lower values
  could be achieved if one considers instead a cored DM profile.
Realistic mass estimates could only be achieved by detailed mass
  modelling once \tol's detailed kinematic information becomes
  available.

New IFU observations are needed to confirm \tol's kinematics.
This could be done with the Multi Unit Spectroscopic Explorer (MUSE) 
\citep{2014Msngr.157...13B}, the Gemini Multi-Object Spectrographs (GMOS) 
\citep{2004PASP..116..425H} or the Fibre Large Array Multi Element
Spectrograph (FLAMES) \citep{2002Msngr.110....1P} as they have the spatial
resolution ($\sim0.5^{\prime\prime}$), spectral coverage and field
of view required to produce H$\alpha$ velocity maps to perform the
kind of study presented by \citet{2015A&A...577A..21C} on CDGs or by
\citet{Herenz16} on nearby \lya\ emitting galaxies.

\section{Conclusions}
\label{sec:conclusions}

In this paper we presented two kinematic models that independently
reproduce the so far unexplained observational features of \tol's
\lya emission line.
One model is based on a multiphase ISM with random clump motions and
the other on gas bulk solid body rotation. 
It is the first time that an observed \lya profile can be 
reproduced by either of these two kinematic conditions.
Our findings highlight the importance of including multiphase and/or
rotation conditions as kinematic features to model the \lya line.

In this particular case, we prefer the multiphase model because it has kinematic
conditions similar to other compact dwarf galaxies observed
with integral field unit spectroscopy, while the
rotational model produces rotational velocities too high for a dwarf
galaxy.  
New IFU observations are the only way to be certain about the detailed
kinematic structure in \tol.  

All in all, the mere existence of a broad SLAE is interesting.  
\tol's line shape is different to others and seems to reveal an
unexpected kinematic structure.
A confirmation of our results by new observations would support the multiphase model
as an element to be included in the study of \lya emitting
galaxies, consolidating the possibility to use the \lya line to
  constrain different models of star formation and feedback in the
  first generation of galaxies. 
 
\section*{Acknowledgments}
We thank the referee for detailed comments and suggestions that
improved the paper. 
We thank J. M. Mas-Hesse for providing us with \tol's observational
\lya data \citep{mashesse03} in electronic format. This data was
used to prepare Figure 1.
J. E. F-R. acknowledges the support of J. L. Guerra in the last
writing stages of this paper.
We are thankful to the community developing and maintaining open source
packages fundamental to our work: numpy \& scipy \citep{numpy}, the
Jupyter notebook \citep{IPython,jupyter}, matplotlib
\citep{matplotlib}, astropy \citep{astropy},  scikit-learn
\citep{scikit-learn}, pandas \citep{pandas} and daft
(\url{http://daft-pgm.org/}).  

\bibliographystyle{mnras}

%\bibliography{references}

\begin{thebibliography}{}
\makeatletter
\relax
\def\mn@urlcharsother{\let\do\@makeother \do\$\do\&\do\#\do\^\do\_\do\%\do\~}
\def\mn@doi{\begingroup\mn@urlcharsother \@ifnextchar [ {\mn@doi@}
  {\mn@doi@[]}}
\def\mn@doi@[#1]#2{\def\@tempa{#1}\ifx\@tempa\@empty \href
  {http://dx.doi.org/#2} {doi:#2}\else \href {http://dx.doi.org/#2} {#1}\fi
  \endgroup}
\def\mn@eprint#1#2{\mn@eprint@#1:#2::\@nil}
\def\mn@eprint@arXiv#1{\href {http://arxiv.org/abs/#1} {{\tt arXiv:#1}}}
\def\mn@eprint@dblp#1{\href {http://dblp.uni-trier.de/rec/bibtex/#1.xml}
  {dblp:#1}}
\def\mn@eprint@#1:#2:#3:#4\@nil{\def\@tempa {#1}\def\@tempb {#2}\def\@tempc
  {#3}\ifx \@tempc \@empty \let \@tempc \@tempb \let \@tempb \@tempa \fi \ifx
  \@tempb \@empty \def\@tempb {arXiv}\fi \@ifundefined
  {mn@eprint@\@tempb}{\@tempb:\@tempc}{\expandafter \expandafter \csname
  mn@eprint@\@tempb\endcsname \expandafter{\@tempc}}}

\bibitem[\protect\citeauthoryear{{Astropy Collaboration} et~al.,}{{Astropy
  Collaboration} et~al.}{2013}]{astropy}
{Astropy Collaboration} et~al., 2013, \mn@doi [\aap]
  {10.1051/0004-6361/201322068}, \href
  {http://adsabs.harvard.edu/abs/2013A%26A...558A..33A} {558, A33}

\bibitem[\protect\citeauthoryear{{Bacon} et~al.,}{{Bacon}
  et~al.}{2014}]{2014Msngr.157...13B}
{Bacon} R.,  et~al., 2014, The Messenger, \href
  {http://adsabs.harvard.edu/abs/2014Msngr.157...13B} {157, 13}

\bibitem[\protect\citeauthoryear{{Begum}, {Chengalur}  \&
  {Karachentsev}}{{Begum} et~al.}{2005}]{2005A&A...433L...1B}
{Begum} A.,  {Chengalur} J.~N.,   {Karachentsev} I.~D.,  2005, \mn@doi [\aap]
  {10.1051/0004-6361:200500026}, \href
  {http://adsabs.harvard.edu/abs/2005A%26A...433L...1B} {433, L1}

\bibitem[\protect\citeauthoryear{{Cair{\'o}s} \&
  {Gonz{\'a}lez-P{\'e}rez}}{{Cair{\'o}s} \&
  {Gonz{\'a}lez-P{\'e}rez}}{2017a}]{2017arXiv170809407C}
{Cair{\'o}s} L.~M.,  {Gonz{\'a}lez-P{\'e}rez} J.~N.,  2017a, preprint, \href
  {http://adsabs.harvard.edu/abs/2017arXiv170809407C} {} (\mn@eprint {arXiv}
  {1708.09407})

\bibitem[\protect\citeauthoryear{{Cair{\'o}s} \&
  {Gonz{\'a}lez-P{\'e}rez}}{{Cair{\'o}s} \&
  {Gonz{\'a}lez-P{\'e}rez}}{2017b}]{2017A&A...600A.125C}
{Cair{\'o}s} L.~M.,  {Gonz{\'a}lez-P{\'e}rez} J.~N.,  2017b, \mn@doi [\aap]
  {10.1051/0004-6361/201629681}, \href
  {http://adsabs.harvard.edu/abs/2017A%26A...600A.125C} {600, A125}

\bibitem[\protect\citeauthoryear{{Cair{\'o}s}, {Caon}  \&
  {Weilbacher}}{{Cair{\'o}s} et~al.}{2015}]{2015A&A...577A..21C}
{Cair{\'o}s} L.~M.,  {Caon} N.,   {Weilbacher} P.~M.,  2015, \mn@doi [\aap]
  {10.1051/0004-6361/201322518}, \href
  {http://adsabs.harvard.edu/abs/2015A%26A...577A..21C} {577, A21}

\bibitem[\protect\citeauthoryear{{Dey}, {Spinrad}, {Stern}, {Graham}  \&
  {Chaffee}}{{Dey} et~al.}{1998}]{1998ApJ...498L..93D}
{Dey} A.,  {Spinrad} H.,  {Stern} D.,  {Graham} J.~R.,   {Chaffee} F.~H.,
  1998, \mn@doi [\apjl] {10.1086/311331}, \href
  {http://adsabs.harvard.edu/abs/1998ApJ...498L..93D} {498, L93}

\bibitem[\protect\citeauthoryear{{Dijkstra}}{{Dijkstra}}{2014}]{2014PASA...31...40D}
{Dijkstra} M.,  2014, \mn@doi [\pasa] {10.1017/pasa.2014.33}, \href
  {http://adsabs.harvard.edu/abs/2014PASA...31...40D} {31, e040}

\bibitem[\protect\citeauthoryear{{Dijkstra} \& {Kramer}}{{Dijkstra} \&
  {Kramer}}{2012}]{2012MNRAS.424.1672D}
{Dijkstra} M.,  {Kramer} R.,  2012, \mn@doi [\mnras]
  {10.1111/j.1365-2966.2012.21131.x}, \href
  {http://adsabs.harvard.edu/abs/2012MNRAS.424.1672D} {424, 1672}

\bibitem[\protect\citeauthoryear{{Dijkstra}, {Lidz}  \& {Wyithe}}{{Dijkstra}
  et~al.}{2007}]{2007MNRAS.377.1175D}
{Dijkstra} M.,  {Lidz} A.,   {Wyithe} J.~S.~B.,  2007, \mn@doi [\mnras]
  {10.1111/j.1365-2966.2007.11666.x}, \href
  {http://adsabs.harvard.edu/abs/2007MNRAS.377.1175D} {377, 1175}

\bibitem[\protect\citeauthoryear{{Engelbracht}, {Rieke}, {Gordon}, {Smith},
  {Werner}, {Moustakas}, {Willmer}  \& {Vanzi}}{{Engelbracht}
  et~al.}{2008}]{2008ApJ...678..804E}
{Engelbracht} C.~W.,  {Rieke} G.~H.,  {Gordon} K.~D.,  {Smith} J.-D.~T.,
  {Werner} M.~W.,  {Moustakas} J.,  {Willmer} C.~N.~A.,   {Vanzi} L.,  2008,
  \mn@doi [\apj] {10.1086/529513}, \href
  {http://adsabs.harvard.edu/abs/2008ApJ...678..804E} {678, 804}

\bibitem[\protect\citeauthoryear{{Erb} et~al.,}{{Erb} et~al.}{2014}]{Erb14}
{Erb} D.~K.,  et~al., 2014, \mn@doi [\apj] {10.1088/0004-637X/795/1/33}, \href
  {http://adsabs.harvard.edu/abs/2014ApJ...795...33E} {795, 33}

\bibitem[\protect\citeauthoryear{{Eskew}, {Zaritsky}  \& {Meidt}}{{Eskew}
  et~al.}{2012}]{2012AJ....143..139E}
{Eskew} M.,  {Zaritsky} D.,   {Meidt} S.,  2012, \mn@doi [\aj]
  {10.1088/0004-6256/143/6/139}, \href
  {http://adsabs.harvard.edu/abs/2012AJ....143..139E} {143, 139}

\bibitem[\protect\citeauthoryear{{Foreman-Mackey}, {Hogg}, {Lang}  \&
  {Goodman}}{{Foreman-Mackey} et~al.}{2013}]{2013PASP..125..306F}
{Foreman-Mackey} D.,  {Hogg} D.~W.,  {Lang} D.,   {Goodman} J.,  2013, \mn@doi
  [\pasp] {10.1086/670067}, \href
  {http://adsabs.harvard.edu/abs/2013PASP..125..306F} {125, 306}

\bibitem[\protect\citeauthoryear{{Forero-Romero}, {Yepes}, {Gottl{\"o}ber},
  {Knollmann}, {Cuesta}  \& {Prada}}{{Forero-Romero} et~al.}{2011}]{CLARA}
{Forero-Romero} J.~E.,  {Yepes} G.,  {Gottl{\"o}ber} S.,  {Knollmann} S.~R.,
  {Cuesta} A.~J.,   {Prada} F.,  2011, \mn@doi [\mnras]
  {10.1111/j.1365-2966.2011.18983.x}, \href
  {http://adsabs.harvard.edu/abs/2011MNRAS.415.3666F} {415, 3666}

\bibitem[\protect\citeauthoryear{{Fricke}, {Izotov}, {Papaderos}, {Guseva}  \&
  {Thuan}}{{Fricke} et~al.}{2001}]{2001AJ....121..169F}
{Fricke} K.~J.,  {Izotov} Y.~I.,  {Papaderos} P.,  {Guseva} N.~G.,   {Thuan}
  T.~X.,  2001, \mn@doi [\aj] {10.1086/318016}, \href
  {http://adsabs.harvard.edu/abs/2001AJ....121..169F} {121, 169}

\bibitem[\protect\citeauthoryear{{Garavito-Camargo}, {Forero-Romero}  \&
  {Dijkstra}}{{Garavito-Camargo} et~al.}{2014}]{GaravitoCamargo2014}
{Garavito-Camargo} J.~N.,  {Forero-Romero} J.~E.,   {Dijkstra} M.,  2014,
  \mn@doi [\apj] {10.1088/0004-637X/795/2/120}, \href
  {http://adsabs.harvard.edu/abs/2014ApJ...795..120G} {795, 120}

\bibitem[\protect\citeauthoryear{{Gavil{\'a}n}, {Ascasibar}, {Moll{\'a}}  \&
  {D{\'{\i}}az}}{{Gavil{\'a}n} et~al.}{2013}]{2013MNRAS.434.2491G}
{Gavil{\'a}n} M.,  {Ascasibar} Y.,  {Moll{\'a}} M.,   {D{\'{\i}}az} {\'A}.~I.,
  2013, \mn@doi [\mnras] {10.1093/mnras/stt1186}, \href
  {http://adsabs.harvard.edu/abs/2013MNRAS.434.2491G} {434, 2491}

\bibitem[\protect\citeauthoryear{{Gawiser} et~al.,}{{Gawiser}
  et~al.}{2007}]{2007ApJ...671..278G}
{Gawiser} E.,  et~al., 2007, \mn@doi [\apj] {10.1086/522955}, \href
  {http://adsabs.harvard.edu/abs/2007ApJ...671..278G} {671, 278}

\bibitem[\protect\citeauthoryear{Goodman \& Weare}{Goodman \&
  Weare}{2010}]{goodman2010ensemble}
Goodman J.,  Weare J.,  2010, Communications in applied mathematics and
  computational science, 5, 65

\bibitem[\protect\citeauthoryear{{Greig}, {Komatsu}  \& {Wyithe}}{{Greig}
  et~al.}{2013}]{2013MNRAS.431.1777G}
{Greig} B.,  {Komatsu} E.,   {Wyithe} J.~S.~B.,  2013, \mn@doi [\mnras]
  {10.1093/mnras/stt292}, \href
  {http://adsabs.harvard.edu/abs/2013MNRAS.431.1777G} {431, 1777}

\bibitem[\protect\citeauthoryear{{Gronke} \& {Dijkstra}}{{Gronke} \&
  {Dijkstra}}{2016}]{Gronke2016}
{Gronke} M.,  {Dijkstra} M.,  2016, \mn@doi [\apj]
  {10.3847/0004-637X/826/1/14}, \href
  {http://adsabs.harvard.edu/abs/2016ApJ...826...14G} {826, 14}

\bibitem[\protect\citeauthoryear{{Gronke}, {Bull}  \& {Dijkstra}}{{Gronke}
  et~al.}{2015}]{2015ApJ...812..123G}
{Gronke} M.,  {Bull} P.,   {Dijkstra} M.,  2015, \mn@doi [\apj]
  {10.1088/0004-637X/812/2/123}, \href
  {http://adsabs.harvard.edu/abs/2015ApJ...812..123G} {812, 123}

\bibitem[\protect\citeauthoryear{{Harrington}}{{Harrington}}{1973}]{Harrington73}
{Harrington} J.~P.,  1973, \mnras, \href
  {http://adsabs.harvard.edu/abs/1973MNRAS.162...43H} {162, 43}

\bibitem[\protect\citeauthoryear{{Hayashino} et~al.,}{{Hayashino}
  et~al.}{2004}]{2004AJ....128.2073H}
{Hayashino} T.,  et~al., 2004, \mn@doi [\aj] {10.1086/424935}, \href
  {http://adsabs.harvard.edu/abs/2004AJ....128.2073H} {128, 2073}

\bibitem[\protect\citeauthoryear{{Hayes}}{{Hayes}}{2015}]{Hayes15}
{Hayes} M.,  2015, \mn@doi [\pasa] {10.1017/pasa.2015.25}, \href
  {http://adsabs.harvard.edu/abs/2015PASA...32...27H} {32, e027}

\bibitem[\protect\citeauthoryear{{Herenz} et~al.,}{{Herenz}
  et~al.}{2016}]{Herenz16}
{Herenz} E.~C.,  et~al., 2016, \mn@doi [\aap] {10.1051/0004-6361/201527373},
  \href {http://adsabs.harvard.edu/abs/2016A%26A...587A..78H} {587, A78}

\bibitem[\protect\citeauthoryear{{Hook}, {J{\o}rgensen}, {Allington-Smith},
  {Davies}, {Metcalfe}, {Murowinski}  \& {Crampton}}{{Hook}
  et~al.}{2004}]{2004PASP..116..425H}
{Hook} I.~M.,  {J{\o}rgensen} I.,  {Allington-Smith} J.~R.,  {Davies} R.~L.,
  {Metcalfe} N.,  {Murowinski} R.~G.,   {Crampton} D.,  2004, \mn@doi [\pasp]
  {10.1086/383624}, \href {http://adsabs.harvard.edu/abs/2004PASP..116..425H}
  {116, 425}

\bibitem[\protect\citeauthoryear{{Hummer} \& {Storey}}{{Hummer} \&
  {Storey}}{1987}]{Hummer1987}
{Hummer} D.~G.,  {Storey} P.~J.,  1987, \mn@doi [\mnras]
  {10.1093/mnras/224.3.801}, \href
  {http://adsabs.harvard.edu/abs/1987MNRAS.224..801H} {224, 801}

\bibitem[\protect\citeauthoryear{{Hunter}}{{Hunter}}{2007}]{matplotlib}
{Hunter} J.~D.,  2007, \mn@doi [Computing in Science and Engineering]
  {10.1109/MCSE.2007.55}, \href
  {http://adsabs.harvard.edu/abs/2007CSE.....9...90H} {9, 90}

\bibitem[\protect\citeauthoryear{{Izotov}, {Papaderos}, {Guseva}, {Fricke}  \&
  {Thuan}}{{Izotov} et~al.}{2004}]{Izotov04}
{Izotov} Y.~I.,  {Papaderos} P.,  {Guseva} N.~G.,  {Fricke} K.~J.,   {Thuan}
  T.~X.,  2004, \mn@doi [\aap] {10.1051/0004-6361:20035847}, \href
  {http://adsabs.harvard.edu/abs/2004A%26A...421..539I} {421, 539}

\bibitem[\protect\citeauthoryear{James, Witten  \& Hastie}{James
  et~al.}{2014}]{james2014introduction}
James G.,  Witten D.,   Hastie T.,  2014, An Introduction to Statistical
  Learning: With Applications in R.

\bibitem[\protect\citeauthoryear{{Kennicutt}}{{Kennicutt}}{1998}]{Kennicutt98}
{Kennicutt} Jr. R.~C.,  1998, \mn@doi [\araa] {10.1146/annurev.astro.36.1.189},
  \href {http://adsabs.harvard.edu/abs/1998ARA%26A..36..189K} {36, 189}

\bibitem[\protect\citeauthoryear{Kluyver et~al.,}{Kluyver
  et~al.}{2016}]{jupyter}
Kluyver T.,  et~al., 2016, in ELPUB. pp 87--90

\bibitem[\protect\citeauthoryear{{Kova{\v c}}, {Somerville}, {Rhoads},
  {Malhotra}  \& {Wang}}{{Kova{\v c}} et~al.}{2007}]{2007ApJ...668...15K}
{Kova{\v c}} K.,  {Somerville} R.~S.,  {Rhoads} J.~E.,  {Malhotra} S.,   {Wang}
  J.,  2007, \mn@doi [\apj] {10.1086/520668}, \href
  {http://adsabs.harvard.edu/abs/2007ApJ...668...15K} {668, 15}

\bibitem[\protect\citeauthoryear{{Laursen}, {Duval}  \& {{\"O}stlin}}{{Laursen}
  et~al.}{2013}]{Laursen2013ApJ...766..124L}
{Laursen} P.,  {Duval} F.,   {{\"O}stlin} G.,  2013, \mn@doi [\apj]
  {10.1088/0004-637X/766/2/124}, \href
  {http://adsabs.harvard.edu/abs/2013ApJ...766..124L} {766, 124}

\bibitem[\protect\citeauthoryear{{Le F{\`e}vre} et~al.,}{{Le F{\`e}vre}
  et~al.}{2003}]{2003SPIE.4841.1670L}
{Le F{\`e}vre} O.,  et~al., 2003, in {Iye} M.,  {Moorwood} A.~F.~M.,  eds,
  \procspie Vol. 4841, Instrument Design and Performance for Optical/Infrared
  Ground-based Telescopes. pp 1670--1681, \mn@doi{10.1117/12.460959}

\bibitem[\protect\citeauthoryear{{Loeb} \& {Rybicki}}{{Loeb} \&
  {Rybicki}}{1999}]{LoebRybicki}
{Loeb} A.,  {Rybicki} G.~B.,  1999, \mn@doi [\apj] {10.1086/307844}, \href
  {http://adsabs.harvard.edu/abs/1999ApJ...524..527L} {524, 527}

\bibitem[\protect\citeauthoryear{{Martin}}{{Martin}}{1998}]{1998ApJ...506..222M}
{Martin} C.~L.,  1998, \mn@doi [\apj] {10.1086/306219}, \href
  {http://adsabs.harvard.edu/abs/1998ApJ...506..222M} {506, 222}

\bibitem[\protect\citeauthoryear{{Mas-Hesse}, {Kunth}, {Tenorio-Tagle},
  {Leitherer}, {Terlevich}  \& {Terlevich}}{{Mas-Hesse}
  et~al.}{2003}]{mashesse03}
{Mas-Hesse} J.~M.,  {Kunth} D.,  {Tenorio-Tagle} G.,  {Leitherer} C.,
  {Terlevich} R.~J.,   {Terlevich} E.,  2003, \mn@doi [\apj] {10.1086/379116},
  \href {http://adsabs.harvard.edu/abs/2003ApJ...598..858M} {598, 858}

\bibitem[\protect\citeauthoryear{{McKee} \& {Ostriker}}{{McKee} \&
  {Ostriker}}{1977}]{1977ApJ...218..148M}
{McKee} C.~F.,  {Ostriker} J.~P.,  1977, \mn@doi [\apj] {10.1086/155667}, \href
  {http://esoads.eso.org/abs/1977ApJ...218..148M} {218, 148}

\bibitem[\protect\citeauthoryear{McKinney et~al.}{McKinney
  et~al.}{2010}]{pandas}
McKinney W.,  et~al., 2010, in Proceedings of the 9th Python in Science
  Conference. pp 51--56

\bibitem[\protect\citeauthoryear{{Mej{\'{\i}}a-Restrepo} \&
  {Forero-Romero}}{{Mej{\'{\i}}a-Restrepo} \&
  {Forero-Romero}}{2016}]{2016ApJ...828....5M}
{Mej{\'{\i}}a-Restrepo} J.~E.,  {Forero-Romero} J.~E.,  2016, \mn@doi [\apj]
  {10.3847/0004-637X/828/1/5}, \href
  {http://adsabs.harvard.edu/abs/2016ApJ...828....5M} {828, 5}

\bibitem[\protect\citeauthoryear{{Neufeld}}{{Neufeld}}{1991}]{1991ApJ...370L..85N}
{Neufeld} D.~A.,  1991, \mn@doi [\apjl] {10.1086/185983}, \href
  {http://adsabs.harvard.edu/abs/1991ApJ...370L..85N} {370, L85}

\bibitem[\protect\citeauthoryear{{Noeske}, {Papaderos}, {Cair{\'o}s}  \&
  {Fricke}}{{Noeske} et~al.}{2003}]{2003A&A...410..481N}
{Noeske} K.~G.,  {Papaderos} P.,  {Cair{\'o}s} L.~M.,   {Fricke} K.~J.,  2003,
  \mn@doi [\aap] {10.1051/0004-6361:20031147}, \href
  {http://adsabs.harvard.edu/abs/2003A%26A...410..481N} {410, 481}

\bibitem[\protect\citeauthoryear{{Orsi}, {Lacey}, {Baugh}  \& {Infante}}{{Orsi}
  et~al.}{2008}]{2008MNRAS.391.1589O}
{Orsi} A.,  {Lacey} C.~G.,  {Baugh} C.~M.,   {Infante} L.,  2008, \mn@doi
  [\mnras] {10.1111/j.1365-2966.2008.14010.x}, \href
  {http://adsabs.harvard.edu/abs/2008MNRAS.391.1589O} {391, 1589}

\bibitem[\protect\citeauthoryear{{Orsi}, {Lacey}  \& {Baugh}}{{Orsi}
  et~al.}{2012}]{Orsi12}
{Orsi} A.,  {Lacey} C.~G.,   {Baugh} C.~M.,  2012, \mn@doi [\mnras]
  {10.1111/j.1365-2966.2012.21396.x}, \href
  {http://adsabs.harvard.edu/abs/2012MNRAS.425...87O} {425, 87}

\bibitem[\protect\citeauthoryear{{{\"O}stlin} et~al.,}{{{\"O}stlin}
  et~al.}{2014}]{LARS}
{{\"O}stlin} G.,  et~al., 2014, \mn@doi [\apj] {10.1088/0004-637X/797/1/11},
  \href {http://adsabs.harvard.edu/abs/2014ApJ...797...11O} {797, 11}

\bibitem[\protect\citeauthoryear{{Ott}, {Walter}  \& {Brinks}}{{Ott}
  et~al.}{2005}]{2005MNRAS.358.1453O}
{Ott} J.,  {Walter} F.,   {Brinks} E.,  2005, \mn@doi [\mnras]
  {10.1111/j.1365-2966.2005.08863.x}, \href
  {http://adsabs.harvard.edu/abs/2005MNRAS.358.1453O} {358, 1453}

\bibitem[\protect\citeauthoryear{{Padilla}, {Christlein}, {Gawiser},
  {Gonz{\'a}lez}, {Guaita}  \& {Infante}}{{Padilla}
  et~al.}{2010}]{2010MNRAS.409..184P}
{Padilla} N.~D.,  {Christlein} D.,  {Gawiser} E.,  {Gonz{\'a}lez} R.~E.,
  {Guaita} L.,   {Infante} L.,  2010, \mn@doi [\mnras]
  {10.1111/j.1365-2966.2010.17317.x}, \href
  {http://adsabs.harvard.edu/abs/2010MNRAS.409..184P} {409, 184}

\bibitem[\protect\citeauthoryear{{Partridge} \& {Peebles}}{{Partridge} \&
  {Peebles}}{1967}]{PartridgePeebles}
{Partridge} R.~B.,  {Peebles} P.~J.~E.,  1967, \mn@doi [\apj] {10.1086/149079},
  \href {http://adsabs.harvard.edu/abs/1967ApJ...147..868P} {147, 868}

\bibitem[\protect\citeauthoryear{{Pasquini} et~al.,}{{Pasquini}
  et~al.}{2002}]{2002Msngr.110....1P}
{Pasquini} L.,  et~al., 2002, The Messenger, \href
  {http://esoads.eso.org/abs/2002Msngr.110....1P} {110, 1}

\bibitem[\protect\citeauthoryear{Pedregosa et~al.,}{Pedregosa
  et~al.}{2011}]{scikit-learn}
Pedregosa F.,  et~al., 2011, Journal of Machine Learning Research, 12, 2825

\bibitem[\protect\citeauthoryear{P\'erez \& Granger}{P\'erez \&
  Granger}{2007}]{IPython}
P\'erez F.,  Granger B.~E.,  2007, \mn@doi [Computing in Science and
  Engineering] {10.1109/MCSE.2007.53}, 9, 21

\bibitem[\protect\citeauthoryear{{Poveda-Ruiz}, {Forero-Romero}  \&
  {Mu{\~n}oz-Cuartas}}{{Poveda-Ruiz} et~al.}{2016}]{2016ApJ...832..169P}
{Poveda-Ruiz} C.~N.,  {Forero-Romero} J.~E.,   {Mu{\~n}oz-Cuartas} J.~C.,
  2016, \mn@doi [\apj] {10.3847/0004-637X/832/2/169}, \href
  {http://adsabs.harvard.edu/abs/2016ApJ...832..169P} {832, 169}

\bibitem[\protect\citeauthoryear{{Prada}, {Klypin}, {Cuesta}, {Betancort-Rijo}
  \& {Primack}}{{Prada} et~al.}{2012}]{2012MNRAS.423.3018P}
{Prada} F.,  {Klypin} A.~A.,  {Cuesta} A.~J.,  {Betancort-Rijo} J.~E.,
  {Primack} J.,  2012, \mn@doi [\mnras] {10.1111/j.1365-2966.2012.21007.x},
  \href {http://adsabs.harvard.edu/abs/2012MNRAS.423.3018P} {423, 3018}

\bibitem[\protect\citeauthoryear{{Pustilnik} \& {Martin}}{{Pustilnik} \&
  {Martin}}{2007}]{pustilnikmartin07}
{Pustilnik} S.~A.,  {Martin} J.-M.,  2007, \mn@doi [\aap]
  {10.1051/0004-6361:20066137}, \href
  {http://cdsads.u-strasbg.fr/abs/2007A%26A...464..859P} {464, 859}

\bibitem[\protect\citeauthoryear{{Rivera-Thorsen} et~al.,}{{Rivera-Thorsen}
  et~al.}{2015}]{2015ApJ...805...14R}
{Rivera-Thorsen} T.~E.,  et~al., 2015, \mn@doi [\apj]
  {10.1088/0004-637X/805/1/14}, \href
  {http://adsabs.harvard.edu/abs/2015ApJ...805...14R} {805, 14}

\bibitem[\protect\citeauthoryear{{Steidel}, {Erb}, {Shapley}, {Pettini},
  {Reddy}, {Bogosavljevi{\'c}}, {Rudie}  \& {Rakic}}{{Steidel}
  et~al.}{2010}]{2010ApJ...717..289S}
{Steidel} C.~C.,  {Erb} D.~K.,  {Shapley} A.~E.,  {Pettini} M.,  {Reddy} N.,
  {Bogosavljevi{\'c}} M.,  {Rudie} G.~C.,   {Rakic} O.,  2010, \mn@doi [\apj]
  {10.1088/0004-637X/717/1/289}, \href
  {http://adsabs.harvard.edu/abs/2010ApJ...717..289S} {717, 289}

\bibitem[\protect\citeauthoryear{{Swaters}, {Sancisi}, {van Albada}  \& {van
  der Hulst}}{{Swaters} et~al.}{2009}]{2009A&A...493..871S}
{Swaters} R.~A.,  {Sancisi} R.,  {van Albada} T.~S.,   {van der Hulst} J.~M.,
  2009, \mn@doi [\aap] {10.1051/0004-6361:200810516}, \href
  {http://adsabs.harvard.edu/abs/2009A%26A...493..871S} {493, 871}

\bibitem[\protect\citeauthoryear{{Tasitsiomi}}{{Tasitsiomi}}{2006}]{2006ApJ...645..792T}
{Tasitsiomi} A.,  2006, \mn@doi [\apj] {10.1086/504460}, \href
  {http://adsabs.harvard.edu/abs/2006ApJ...645..792T} {645, 792}

\bibitem[\protect\citeauthoryear{{Tassis}, {Kravtsov}  \& {Gnedin}}{{Tassis}
  et~al.}{2008}]{2008ApJ...672..888T}
{Tassis} K.,  {Kravtsov} A.~V.,   {Gnedin} N.~Y.,  2008, \mn@doi [\apj]
  {10.1086/523880}, \href {http://adsabs.harvard.edu/abs/2008ApJ...672..888T}
  {672, 888}

\bibitem[\protect\citeauthoryear{{Thuan} \& {Izotov}}{{Thuan} \&
  {Izotov}}{1997}]{Thuan97}
{Thuan} T.~X.,  {Izotov} Y.~I.,  1997, \apj, \href
  {http://adsabs.harvard.edu/abs/1997ApJ...489..623T} {489, 623}

\bibitem[\protect\citeauthoryear{{Tollerud}, {Bullock}, {Graves}  \&
  {Wolf}}{{Tollerud} et~al.}{2011}]{2011ApJ...726..108T}
{Tollerud} E.~J.,  {Bullock} J.~S.,  {Graves} G.~J.,   {Wolf} J.,  2011,
  \mn@doi [\apj] {10.1088/0004-637X/726/2/108}, \href
  {http://adsabs.harvard.edu/abs/2011ApJ...726..108T} {726, 108}

\bibitem[\protect\citeauthoryear{{Trainor}, {Strom}, {Steidel}  \&
  {Rudie}}{{Trainor} et~al.}{2016}]{Trainor16}
{Trainor} R.~F.,  {Strom} A.~L.,  {Steidel} C.~C.,   {Rudie} G.~C.,  2016,
  \mn@doi [\apj] {10.3847/0004-637X/832/2/171}, \href
  {http://adsabs.harvard.edu/abs/2016ApJ...832..171T} {832, 171}

\bibitem[\protect\citeauthoryear{{Verhamme}, {Schaerer}  \&
  {Maselli}}{{Verhamme} et~al.}{2006}]{2006A&A...460..397V}
{Verhamme} A.,  {Schaerer} D.,   {Maselli} A.,  2006, \mn@doi [\aap]
  {10.1051/0004-6361:20065554}, \href
  {http://adsabs.harvard.edu/abs/2006A%26A...460..397V} {460, 397}

\bibitem[\protect\citeauthoryear{{Verhamme}, {Orlitov{\'a}}, {Schaerer}  \&
  {Hayes}}{{Verhamme} et~al.}{2015}]{2015A&A...578A...7V}
{Verhamme} A.,  {Orlitov{\'a}} I.,  {Schaerer} D.,   {Hayes} M.,  2015, \mn@doi
  [\aap] {10.1051/0004-6361/201423978}, \href
  {http://adsabs.harvard.edu/abs/2015A%26A...578A...7V} {578, A7}

\bibitem[\protect\citeauthoryear{Walt, Colbert  \& Varoquaux}{Walt
  et~al.}{2011}]{numpy}
Walt S. v.~d.,  Colbert S.~C.,   Varoquaux G.,  2011, Computing in Science \&
  Engineering, 13, 22

\bibitem[\protect\citeauthoryear{{Yamada}, {Matsuda}, {Kousai}, {Hayashino},
  {Morimoto}  \& {Umemura}}{{Yamada} et~al.}{2012}]{2012ApJ...751...29Y}
{Yamada} T.,  {Matsuda} Y.,  {Kousai} K.,  {Hayashino} T.,  {Morimoto} N.,
  {Umemura} M.,  2012, \mn@doi [\apj] {10.1088/0004-637X/751/1/29}, \href
  {http://adsabs.harvard.edu/abs/2012ApJ...751...29Y} {751, 29}

\makeatother
\end{thebibliography}

\appendix
\section{Random Forest Classification}
\label{appendix}

As a complement to the K-S tests on the multiphase data 
we apply a random forest classification algorithm
\citep{james2014introduction} to find the
more relevant parameters in the model to produce a low $\chi^2$
result.

We divide
the results in two classes: low $\chi^2/d.o.f.<35$ and high
$\chi^2/d.o.f.>35$, that is the limit that divides the best $1\%$ of
the models from the rest. 
The algorithm uses $500$ trees for the classification and a
maximum of $3$ depth levels. 
To check for stability we repeat the computation $10$ times by
randomly subsampling the input data to use $80\%$ of the data as
the training set.

\begin{figure*}
\begin{center}
\includegraphics[width=0.95\textwidth]{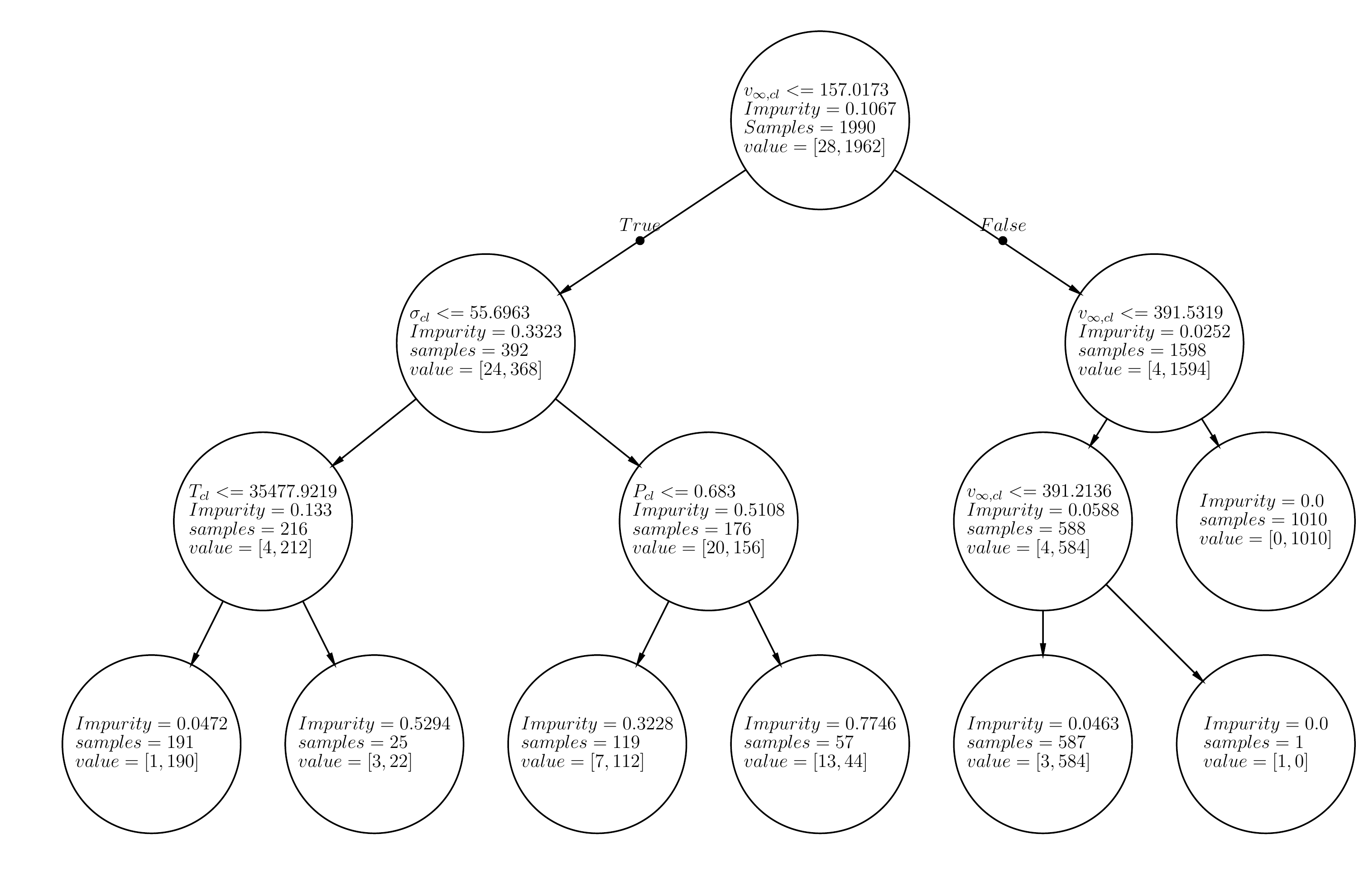}
\caption{Classification tree example. 
The aim is to find the parameters that can be used to separate the
results in two classes: low $\chi^2/d.o.f. < 35$ and high $\chi^2/d.o.f>35$. 
Each box describes the condition over the parameter of interest, the
level of sample impurity, the total number of samples and the value of the
number of samples in each class.  
In this example we randomly sample $80\%$ of the full data set of models to
start with $28$ models in the first class and
$1962$ models in the second class; the best way to increase the
probability to find a result with low $\chi^2/d.o.f.$ ($13$ models,
fourth bottom box, from left to the right)
is having the clump radial velocity, $v_{\infty,{\rm cl}}<157.0$ \kms;
the clump velocity dispersion, $\sigma_{{\rm cl}}>55.6$ \kms, and the
probability to have a \lya photon emitted in a clump, $P_{\rm cl}>0.683$.
The final classification shows that these three are the most relevant
to select the best models.
\label{fig:tree}}
\end{center}
\end{figure*}

Figure \ref{fig:tree} shows as an example the results for a
single classification tree.
The tree starts with $28$ and $1962$ models in the low and high
$\chi^2/d.o.f.$ classes, respectively. 
In this example the best classification
yields $13$ and $44$ models in the low and high $\chi^2/d.o.f.$
classes, respectively, after selecting for $v_{\infty,{\rm cl}}<157.0$ \kms,
$\sigma_{{\rm cl}}>55.6$ \kms and $P_{\rm cl}>0.683$. 

The results of the random forest classifier over $500$ trees yield
that the clump outflow velocity $v_{\infty,\rm{cl}}$, the clump
velocity dispersion $\sigma_{\rm cl}$ and the probability that the
\lya emission comes from the clumps $P_{\rm cl}$, are the most
influential parameters in finding a model with low $\chi^2/d.o.f.$.

\end{document}